\documentstyle[12pt]{article}
\newcommand{\beq}{\begin{equation}}
\newcommand{\eeq}{\end{equation}}
\newcommand{\id}
 {i\kern.06em\hbox{\raise.25ex\hbox{$/$}\kern-.60em$\partial$}}

\newcommand{\bs}{/\kern-.52em b}
\newcommand{\qs}{/\kern-.52em s}

\newcommand{\cL}{{\cal L}}
\newcommand{\cLg}{{\cal L}_{g}}
\newcommand{\p}{\partial}
\newcommand{\dd}
{\kern.06em\hbox{\raise.25ex\hbox{$/$}\kern-.60em$\partial$}}

\date{}
\begin{document}
\title{Generally Covariant Conservative Energy-Momentum
for Gravitational Anyons
\thanks{On leave of absence from Physics Department, Shanghai University, 201800, 
Shanghai, China}}
\author{{Sze-Shiang Feng$^{1,2,3}$,  Yi-Shi Duan$^{4}$}\\
1.{\small {\it High Energy Section, ICTP, Trieste, 34100,Italy}}\\
  E-mail:fengss@ictp.trieste.it\\
2. {\small {\it CCAST(World Lab.) P. O. Box 8730, Beijing, 100080}}\\
3.{\small {\it Department of Modern Physics,  University
                of Science and Technology of China,
                   230026, Hefei, China}}\\e-mail:zhdp@ustc.edu.cn\\
4.{\small {\it Institute of Theoretical Physics, Lanzhou University,
        730000, Lanzhou,China}}\\e-mail: ysduan@lzu.edu.cn\\}
\maketitle
\begin{center}
\begin{minipage}{120mm}
\vskip 1.in
\baselineskip 0.32in
\begin{center}{\bf Abstract}\end{center}
{We obtain a generally covariant conservation law of energy
momentum for
gravitational anyons by the general displacement transform.
The energy-
momentum currents have also superpotentials and are therefore
identically
conserved. It is shown that for Deser's solution and
Cl$\acute{e}$ment's
solution, the energy vanishes. The reasonableness of the
definition
of energy-momentum may be confirmed by the solution for
pure Einstein
gravity which is a limit of vanishing Chern-Simons coupling of
gravitational
anyons.}
\end{minipage}
\end{center}
\vskip 1in
\newpage
\section*{I.Introduction}
\baselineskip 0.32in
\indent Gravity in 1+2 dimensional spacetime has been a
popular subject of
discussion from a decade ago\cite{s1}-\cite{s6}. Though some
of the models
are toy  models, the studies may shed light on the understanding
of not
only quasi-(1+2) dimensional physics (e.g. QHE and high $T_{c}$),
but also
the realistic 1+3 dimensional gravity. In any gauge theory in
odd dimensions,
there exists  a special term, i.e. the Chern-Simons term,
which can be
incorporated into the model Lagrangian. The concept of
gravitational anyons
is a simple non-Abelian generalization of the U(1)
Chern-Simons theory to
non-compact gauge group\cite{s1}\cite{s2}. Using his sloution
to the
linearized field equations, Deser studied the mass and spin of the
gravitational anyons\cite{s4}. The conclusion states that the
gravitational
and inertial quantities are not equal to each other in general and
thus
the equivalence principle is violated. Hence, there exist much
difference
between gravitational anyons and 1+3 Einstein gravity.\\
\indent It seems to us that in order to understand the difference,
we should
have a well-defined definition of gravitational conservative
quantities, or
in other words, we should have generally covariant conservation
laws. In
our previous work\cite{s7}, we have obtained a generally covariant
conservation law of angular-momentum for gravitational anyons.
As suggested
in\cite{s7}, the present paper is to study the generally covariant
conservation law of energy-momentum in the approach proposed
in\cite{s8}.
The paper is arranged as follows. In section II, we give a general
description of the scheme for establishing generally covariant
conservation
laws in general relativity. In section III, we use the general
displacement
transform and the scheme to obtain a generally covariant
conservation law
of energy-momentum. In this section, we will use a first order
Lagrangian
instead of the original one which is in second order. In section
IV, we
calculate the total energy-momentum for Deser's,
Cle$\acute{e}$ment's
solutions, and a solution in the vanishing Chern-Simons coupling
limit.
The last section is devoted to some remarks and discussions.\\
\section*{II. General Scheme for Conservation Laws in General
Relativity}
\indent As in 1+3 Einstein gravity, conservation laws are also
the consequence
of the invariance of the action corresponding to some transforms.
In order
to study the covariant energy-momentum of more complicated systems,
it is
benifecial to discuss conservation laws by Noether theorem in general.
Suppose that the spacetime is of dimension $D=1+d$ and the Lagrangian
is in the first order formalism, i.e.
\beq
I=\int_{G}\cL(\phi^{A}, \p_{\mu}\phi^{A})d^{D}x
\eeq
where $\phi^{A}$ denotes the generic fields. If the action is invariant
under the infinitesimal transforms
\beq
x^{\prime\mu}=x^{\mu}+\delta x^{\mu}\,\,\,\,\,\,\,\,  \phi^{\prime A}
(x^{\prime})=\phi^{A}(x)+\delta\phi^{A}(x)
\eeq
 (it is not required that $\delta\phi^{A}_{\mid\p G}=0$), then following relation
holds\cite{s8}-
\cite{s10}.
\beq
\p_{\mu}(\cL\delta x^{\mu}+\frac{\p\cL}{\p\p_{\mu}\phi^{A}}
\delta_{0}\phi^{A}
)+[\cL]_{\phi^{A}}\delta_{0}\phi^{A}=0
\eeq
where
\beq
[\cL]_{\phi^{A}}=\frac{\p\cL}{\p\phi^{A}}-\p_{\mu}
\frac{\p\cL}{\p\p_{\mu}\phi^{A}}
\eeq
and $\delta_{0}\phi^{A}$ is the Lie variation of $\phi^{A}$
\beq
\delta_{0}\phi^{A}=\phi^{\prime A}(x)-\phi^{A}(x)=
\delta\phi^{A}(x)-\p_{\mu}
\phi^{A}\delta x^{\mu}
\eeq
\indent If $\cL$ is the total Lagrangian of the system,
the field equations of
$\phi^{A}$ is just $[\cL]_{\phi^{A}}=0$. Hence from eq.(3), we
can obtain the
conservation equation corresponding to transform eq.(2)
\beq
\p_{\mu}(\cL\delta x^{\mu}+\frac{\p\cL}{\p\p_{\mu}\phi^{A}}
\delta_{0}
\phi^{A})=0
\eeq
It is important to recognize that if $\cL$ is not the total
Lagrangian
, e.g. the gravitational part $\cL_{g}$, then so long as the
action of
$\cL_{g}$ remains invariant under transform eq.(2), eq.(3) is
still valid
yet eq.(6) is no longer admissible because of
$[\cL_{g}]_{\phi^{A}}\not=0$.\\
\indent Suppose that $\phi^{A}$ denotes the Riemann tensors
$\phi^{A}_{\mu}$
and Riemann scalars $\psi^{A}$ (for gravitational anyons,
they are dreibein
$e^{a}_{\mu}$, SO(1,2) connection $\omega^{a}_{\mu}$, the Lagrangian
multiplier $\lambda^{a}_{\mu}$ and the matter field $\psi^{A}$).
Eq.(3) reads
(suppose that $\cLg$ does not contain $\psi^{A}$)
\beq
\p_{\mu}(\cLg\delta x^{\mu}+\frac{\p\cLg}{\p\p_{\mu}\phi^{A}_{\nu}}
\delta_{0}
\phi^{A}_{\nu})+[\cLg]_{\phi^{A}_{\mu}}\delta_{0}\phi^{A}_{\mu}=0
\eeq
Under transforms eq.(2), the Lie variations are
\beq
\delta_{0}\phi^{A}_{\nu}=-\delta x^{\alpha}_{,\nu}\phi^{A}_{\alpha}
-\phi^{A}_{\nu,\alpha}\delta x^{\alpha}
\eeq
where the dot "," denotes partial derivative. So eq.(7) reads
\beq
\p_{\mu}[\cLg\delta x^{\mu}-\frac{\p\cLg}{\p\p_{\mu}
\phi^{A}_{\lambda}}
(\delta x^{\nu}_{,\lambda}\phi^{A}_{\nu}+\phi^{A}_{\lambda ,\nu}
\delta x^{\nu})]
-[\cLg]_{\phi^{A}_{\lambda}}(\delta x^{\nu}_{,\lambda}\phi^{A}_{\nu}+
\phi^{A}_{\lambda ,\nu}\delta x^{\nu})=0
\eeq
Comparing the coefficients of $\delta x^{\nu},
\delta x^{\nu}_{,\lambda}
$ and $\delta x^{\nu}_{,\mu\lambda}$, we may obtain an identity
\beq
\p_{\lambda}([\cLg]_{\phi^{A}_{\lambda}}\phi^{A}_{\nu})=
[\cLg]_{\phi^{A}
_{\lambda}}\phi^{A}_{\lambda,\nu}
\eeq
Then eq.(9) can be written as
\beq
\p_{\mu}[\cLg\delta x^{\mu}-\frac{\p\cLg}{\p\p_{\mu}
\phi^{A}_{\lambda}}
(\delta x^{\nu}_{,\lambda}\phi^{A}_{\nu}+\phi^{A}_{\lambda ,\nu}
\delta
x^{\nu})-[\cLg]_{\phi^{A}_{\mu}}\phi^{A}_{\nu}\delta x^{\nu}]=0
\eeq
or
\beq
\p_{\mu}[(\cLg\delta^{\mu}_{\nu}-\frac{\p\cLg}{\p\p_{\mu}
\phi^{A}_{\lambda}}
\phi^{A}_{\lambda,\nu}-[\cLg]_{\phi^{A}_{\mu}}\phi^{A}_{\nu})
\delta x^{\nu}
-\frac{\p\cLg}{\p\phi^{A}_{\lambda,\mu}}\phi^{A}_{\nu}\delta
x^{\nu}_{,\lambda}]=0
\eeq
By definition, we introduce
\beq
\tilde{I}^{\mu}_{\nu}=-(\cLg\delta^{\mu}_{\nu}-
\frac{\p\cLg}{\p\p_{\mu}
\phi^{A}_{\lambda}}\phi^{A}_{\lambda,\nu}-
[\cLg]_{\phi^{A}_{\mu}}\phi^{A}_{\nu})
\eeq
\beq
\tilde{Z}^{\lambda\mu}_{\nu}=\frac{\p\cLg}{\p\phi^{A}_{\lambda,\mu}}
\phi^{A}_{\nu}
\eeq
Then eq.(12) gives
\beq
\p_{\mu}(\tilde{I}^{\mu}_{\nu}\delta x^{\nu}+
\tilde{Z}^{\lambda\mu}_{\nu}
\delta x^{\nu}_{,\lambda})=0
\eeq
So by comparing the coefficients of $\delta x^{\nu},
\delta x^{\nu}_{,\mu}$
and $\delta x^{\nu}_{,\mu\lambda}$, we have the following from eq.(15)
\beq
\p_{\mu}\tilde{I}^{\mu}_{\nu}=0
\eeq
\beq
\tilde{I}^{\lambda}_{\nu}=-\p_{\mu}\tilde{Z}^{\lambda\mu}_{\nu}
\,\,\,\,\,\,\,\,
\tilde{Z}^{\mu\lambda}_{\nu}=-\tilde{Z}^{\lambda\mu}_{\nu}
\eeq
Eq.(16)-(17) are fundamental to the establishing of conservation law of
energy-momentum.\\
\section*{III. Conservation Law of Energy-momentum for Gravitational
Anyons}
{\it 3.1. General Displacement Transform}\\
\indent In 1+3 Einstein gravity, a generally covariant conservation law
of energy-momentum was obtained by means of whant we usually call
the {\it general displacement transform}\cite{s8}
\beq
\delta x^{\mu}=e^{\mu}_{a}\epsilon^{a}\,\,\,\,\,\,\,\,\,\,
(\epsilon^{a}
=const.)
\eeq
It really represents an infinitesiaml displacement while
\beq
\delta x^{\mu}=b^{\mu}=const.
\eeq
does not because $x^{\mu}$ can be any coordinates. For instance,
it can be
spherical coordinates, in which case, the resulting conservation
law, if
exists, should be that of angular-momentum other than energy-momentum.
Using
the invariance of the action with respect to eq.(18) and Einstein
equations
, the following general covariant conservation law of energy-momentum
is
obtained
\beq
\nabla_{\mu}(T^{\mu}_{a}+t^{\mu}_{a})=0
\eeq
and it was shown that there exist superpotentials $V^{\mu\nu}_{a}$
\beq
T^{\mu}_{a}+t^{\mu}_{a}=\nabla_{\nu}V^{\mu\nu}_{a}
\eeq
\beq
V^{\mu\nu}_{a}=\frac{c^{4}}{8\pi G}[e^{\mu}_{b}e^{\nu}_{c}
\omega_{a}\,^{bc}
+(e^{\mu}_{a}e^{\nu}_{b}-e^{\mu}_{b}e^{\nu}_{a})\omega^{b}]
\eeq
This definition of energy-momentum has the following main
propertities:\\
1). It is a covariant definition with respect to general coordinate
transforms. But the energy-momentum tensor is not covariant under
local
Lorentz transforms, this is reasonable because of the equivalence
principle.\\
2). For closed system, the total energy-momentum does not depend
on the
choice of Riemann coordinates and transforms in the covariant way
\beq
P_{a}^{\prime}=L_{a}\,^{b}P_{b}
\eeq
under local Lorentz transform $\Lambda^{a}\,_{b}$ which is constant
$L^{a}\,
_{b}$ at spatial infinity.\\
3). For a closed system with static mass center, the total
energy-momentum is
$P_{a}=(Mc,0,0,0)$.\\
4). For a rather concentrated matter system, the gravitational
energy
radiation is\cite{s11}
\beq
-\frac{\p E}{\p t}=\frac{G}{45c^{5}}(\stackrel{\cdots}{D})^{2}
\eeq
5). For Bondi's plane wave, the energy current is\cite{s11}
\beq
t^{\mu}_{0}=(\frac{1}{4\pi}\beta^{\prime 2},
\frac{1}{4\pi}\beta^{\prime 2}, 0,0)
\eeq
6). For the solution of gravitational solitons, we can obtain
finite energy
while the Landau-Lifshitz definition leads to infinite energy
\cite{s12}.\\
7). In Ashtekar's complex formalism of general relativity,
the energy-momentum and angular-momentum constitute a 3-Poincare
algebra
and the energy coincides with the ADM energy\cite{s10}.\\
With these foundations, we next use 1+2 dimensional transform
eq.(18) to
obtain conservative energy-momentum for gravitational anyons.\\
{\it 3.2 The Energy-momentum for Gravitational Anyons}\\
\indent We take the Lagrangian for gravitational anyons to be
in the
first order ($\omega^{a}_{\mu}=\frac{1}{2}\epsilon^{abc}
\omega_{\mu bc},
\omega_{\mu bc}$ is the SO(1,2) connection.)
\beq
\cL=\cLg+\cL_{m}
\eeq
where
\beq
\cLg=-\frac{1}{\kappa}\epsilon^{\mu\nu\alpha}\omega^{a}_{\nu}\p_{\mu}
e_{\alpha a}+\frac{1}{2\kappa}\epsilon^{\mu\nu\alpha}e_{\alpha a}
\epsilon^{abc}\omega_{\mu b}\omega_{\nu c}
+\cL_{c-s}+\epsilon^{\mu\nu\alpha}\lambda^{a}_{\mu}(\p_{\nu}
e_{\alpha a}
+\epsilon_{abc}\omega^{b}_{\nu}e^{c}_{\alpha})
\eeq
\beq
\cL_{c-s}=\frac{1}{2\kappa\mu}\epsilon^{\mu\nu\alpha}(\omega_{\mu a}
\p_{\nu}\omega^{a}_{\alpha}+\frac{1}{3}\epsilon_{abc}\omega^{a}_{\mu}
\omega^{b}_{\nu}\omega^{c}_{\alpha})
\eeq
and $\cL_{m}$ denotes the matter part. The field equations for
$e^{a}_{\mu},
\omega^{a}_{\mu}$ and $\lambda^{a}_{\mu}$ are $[\cL]_{e^{a}_{\mu}}=0,
[\cL]_{\omega^{a}_{\mu}}=0 $ and $[\cL]_{\lambda^{a}_{\mu}}=0$, i.e.
\beq
\frac{1}{2\kappa}\epsilon^{\mu\nu\alpha}\epsilon^{abc}\omega_{\mu b}
\omega_{\nu c}+\epsilon^{\mu\nu\alpha}\lambda^{b}_{\mu}\epsilon_{abc}
\omega^{c}_{\nu}+\frac{1}{\kappa}\epsilon^{\mu\nu\alpha}\p_{\mu}
\omega^{a}_{\nu}+\epsilon^{\mu\nu\alpha}\p_{\mu}\lambda_{\nu}^{a}=
-[\cL_{m}]_{e^{a}_{\alpha}}
\eeq
\beq
\epsilon^{\mu\nu\alpha}(\p_{\nu}e_{\alpha a}+\epsilon_{abc}
\omega^{b}_{\nu}
e^{c}_{\alpha})=0
\eeq
$$
\frac{1}{\kappa}\epsilon^{\nu\mu\alpha}(\p_{\mu}e^{a}_{\alpha}+
\epsilon^{a}\,
_{bc}\omega^{b}_{\mu}e^{c}_{\alpha})+\frac{1}{\kappa\mu}
\epsilon^{\nu\mu\alpha}
(\p_{\mu}\omega^{a}_{\alpha}$$
\beq
+\frac{1}{2}\epsilon^{abc}\omega_{\mu b}\omega_{\alpha c})+
\epsilon^{\nu\mu
\alpha}\lambda_{b \mu}\epsilon^{abc}e_{\alpha c}
=-[\cL_{m}]_{\omega^{a}_{\nu}}
\eeq
Using eq.(30), eq.(31) can be rewritten as
\beq
\frac{1}{\kappa\mu}\epsilon^{\nu\mu\alpha}
(\p_{\mu}\omega^{a}_{\alpha}
+\frac{1}{2}\epsilon^{abc}\omega_{\mu b}\omega_{\alpha c})+
\epsilon^{\nu\mu
\alpha}\lambda_{b \mu}\epsilon^{abc}e_{\alpha c}
=-[\cL_{m}]_{\omega^{a}_{\nu}}
\eeq
These equations are the same as those given in\cite{s13}.\\
\indent From eq.(14)
$$\tilde{Z}^{\lambda\mu}_{\nu}=\frac{\p\cLg}{\p e^{a}
_{\lambda,\mu}}e^{a}_{\nu}
+\frac{\p\cLg}{\p\omega^{a}_{\lambda,\mu}}\omega^{a}_{\nu}
+\frac{\p\cLg}{\p\lambda^{a}_{\lambda,\mu}}\lambda^{a}_{\nu}$$
\beq
=-\frac{1}{\kappa}\epsilon^{\mu\alpha\lambda}\omega_{\alpha a}
e^{a}_{\nu}
+\frac{1}{2\kappa\mu}\epsilon^{\alpha\mu\lambda}\omega_{\alpha a}
\omega^{a}_{\nu}+\epsilon^{\alpha\mu\lambda}\lambda_{\alpha a}
e^{a}_{\nu}
\eeq
For transform eq.(18), eq.(15) implies
\beq
\p_{\mu}(\tilde{I}^{\mu}_{\nu}e^{\nu}_{a}+\tilde{Z}^{\lambda\mu}_{\nu}
e^{\nu}_{a,\lambda})=0
\eeq
Define
\beq
\tilde{I}^{\mu}_{a}=-\p_{\lambda}\tilde{Z}^{\lambda\mu}_{a},\,\,\,\,
\,\,\,\,\,
\tilde{Z}^{\lambda\mu}_{a}=\tilde{Z}^{\lambda\mu}_{\nu}e^{\nu}_{a}
\eeq
we then have
\beq
\p_{\mu}\tilde{I}^{\mu}_{a}=0
\eeq
Since $[\cLg]_{e^{a}_{\mu}}=-[\cL_{m}]_{e^{a}_{\mu}}$, and
$T^{\mu}_{a}
=-\frac{1}{e}[\cL_{m}]_{e^{a}_{\mu}}$, we have from eq.(13)
\beq
\tilde{I}^{\mu}_{\nu}=-(\cLg\delta^{\mu}_{\nu}-
\frac{\p\cLg}{\p e^{a}_{\lambda,
\mu}}e^{a}_{\lambda,\nu}-\frac{\p\cLg}{\p \omega^{a}_{\lambda,
\mu}}\omega^{a}_{\lambda,\nu}
-[\cLg]_{\omega^{a}_{\mu}}\omega^{a}_{\nu}-
[\cLg]_{\lambda^{a}_{\mu}}
\lambda^{a}_{\nu})+eT^{\mu}_{a}e^{a}_{\nu}
\eeq
Define $t^{\mu}_{a}$ by
\beq
\tilde{I}^{\mu}_{\nu}e^{\nu}_{a}+\tilde{Z}^{\lambda\mu}_{\nu}
e^{\nu}_{a,\lambda}=e(T^{\mu}_{a}+t^{\mu}_{a})
\eeq
Then we have
\beq
e(T^{\mu}_{a}+t^{\mu}_{a})=e\nabla_{\lambda}Z^{\lambda\mu}_{a}
\eeq
where $\tilde{Z}^{\lambda\mu}_{a}=eZ^{\lambda\mu}_{a}$. Eq.(39) is the
desired general covariant conservation law of energy-momentum for
gravitational anyons. The total energy-momentum is
\beq
P_{a}=\int e(T^{0}_{a}+t^{0}_{a}) d^{2}x=\int \p_{i}\tilde{Z}^{i0}_{a}
d^{2}x
\eeq
{\it 3.3 The iso(1,2) Algebra}\\
\indent The pure Einstein case is restored by setting
$\mu\rightarrow\infty$
and $\cL_{m}=0$. In this limit, we have $\lambda^{a}_{\mu}=0$
and the
superpotential is simply
\beq
\tilde{Z}^{\mu\nu}_{a}=-\frac{1}{\kappa}\epsilon^{\mu\nu\alpha}
\omega_{\alpha
a}
\eeq and the total energy-momentum is
\beq
P_{a}=\frac{1}{\kappa}\oint_{\p\Sigma}\omega_{a}
\eeq
where $\p\Sigma$ is the spatial infinity which is 1 dimensional.
From the
angular-momentum\cite{s7}
\beq
J_{a}=-\frac{1}{\kappa}\oint_{\p\Sigma}e_{a}=\frac{1}{\kappa}\int
\epsilon^{ij}\epsilon_{abc}\omega^{b}_{i}e^{c}_{j} d^{2}x
\eeq and the Poisson brackets given in \cite{s5}
$$\{\omega^{a}_{i}(x),e^{b}_{j}(y)\}=\kappa\epsilon_{ij}\eta^{ab}
\delta^{2}(x-y)
$$
\beq
\{\omega^{a}_{i}(x),\omega^{b}_{j}(y)\}=\{e^{a}_{i}(x),e^{b}_{j}
(y)\}=0
\eeq
we have
\beq
\{J_{a},J_{b}\}=-\frac{1}{\kappa^2}\{\oint e_{a}(x), \int
\epsilon^{ij}
\epsilon_{bcd}\omega^{c}_{i}e^{d}_{j} d^{2}y\}=\epsilon_{abc}J^{c}
\eeq
\beq
\{P_{a}, P_{b}\}=0
\eeq
\beq
\{J_{a}, P_{b}\}=\{\frac{1}{\kappa}\int \epsilon^{ij}\epsilon_{acd}
\omega^{c}_{i}e^{d}_{j}d^{2}x, \frac{1}{\kappa}\oint\omega_{b}(y)\}
=\epsilon_{abc}P^{c}
\eeq
Thus the ${\it iso}(1,2)$ algebra can be restored.
\section*{IV. Examples}
\indent We now consider the special case that $[\cL_{m}]_{\omega^{a}
_{\mu}}=0$. Using the identities in 3-dim Riemann geometry
$$
R_{\alpha\beta\gamma\delta}=g_{\alpha\gamma}\tilde{R}_{\beta\delta}+
g_{\beta\delta}\tilde{R}_{\alpha\gamma}-g_{\alpha\delta}\tilde{R}
_{\beta\gamma}-g_{\beta\gamma}\tilde{R}_{\alpha\delta}$$
$$\tilde{R}_{\mu\nu}=R_{\mu\nu}-\frac{1}{4}g_{\mu\nu}R
\,\,\,\,\,\,\,\, R^{\alpha\beta}\,_{\gamma\delta}=
-\epsilon^{\alpha\beta\mu}
\epsilon_{\gamma\delta\nu}G^{\nu}_{\mu}\,\,\,\,\,\,\, G^{\mu\nu}=
R^{\mu\nu}
-\frac{1}{2}g^{\mu\nu}R$$
we have
\beq
-\frac{1}{\kappa\mu}\epsilon^{\mu\alpha\beta}(\p_{\alpha}
\omega_{\beta a}
+\frac{1}{2}\epsilon_{abc}\omega^{b}_{\alpha}\omega^{c}_{\beta})=
\frac{1}{\kappa\mu}eG^{\mu}_{a}
\eeq
So
\beq
\lambda^{a}_{\mu}=-\frac{1}{\kappa\mu}\tilde{R}^{a}_{\mu}
\eeq
Substitute into eq.(29), we have
\beq
G^{\mu}_{a}+\frac{1}{\mu}C^{\mu}_{a}=-\kappa T^{\mu}_{a}
\eeq
where $C^{\mu}_{a}$ is the Cotton tensor. For Deser's solution
\cite{s4}.
$$e^{0}_{0}\simeq 1,\enspace
e^{1}_{1}=e^{2}_{2}\simeq \sqrt{\frac{m\kappa^{2}}{\pi}}
ln^{\frac{1}{2}}r$$
 $$e^{0}_{2}\simeq
-\frac{\kappa^{2}}{\mu}\frac{(m+\mu\sigma)}{2\pi}\frac{x}
 {r^{2}},\enspace e^{0}_{1}\simeq
\frac{\kappa^{2}(m+\mu\sigma)}{2\pi\mu}
\frac{y}{r^{2}}$$
\beq
e^{1}_{2}\simeq
-\frac{\kappa^{4}(m+\mu\sigma)^{2}}{4\pi^{2}\mu^{2}\sqrt{\frac
{m\kappa^{2}}{\pi}}}\frac{xy}{r^{4}ln^{1/2}r}
\eeq
 we can obtain the asymptotical behaviour of the
spin connection \begin{equation}
\omega_{\mu ab}\simeq \frac{1}{rf(r)}
\end{equation}
 where$f(r)$ represents some monototically
increasing functions of $r$. Thus we  have
\begin{equation}
\oint_{\partial\Sigma}\omega_{\mu ab}dx^{\mu}=0
\end{equation}
Thus the total energy-momentum vanishes.
For Cl$\acute{e}$ment's \cite{s14}
 self-dual exact solution
\begin{equation}
 ds^{2}=A^{-1}[dt-(\omega_{0}+A)d\theta]^{2}-dr^{2
}-Ad\theta^{2} \end{equation}
$$A=a+ce^{-\mu r}$$
 where $a,c$ are constants and $\mu$ should be
positive since $r\in (0,+\infty)
$. In terms of rectangular coordinates, it takes
the following form
 $$ds^{2}=A^{-1}dt^{2}-2(\omega_{0}A^{-1}+1)(\frac
{x}{r^{2}}dtdy-\frac{y} {r^{2}}dtdx)$$
 $$+\{[A^{-1}(\omega_{0}+A)^{2}-A]\frac{x^{2}}{r^{
4}}-1\}dy^{2}+\{[A^{-1}
(\omega_{0}+A)^{2}-A]\frac{y^{2}}{r^{4}}-1\}dx^{2}$$
\begin{equation}
-[A^{-1}(\omega_{0}+A)^{2}-A]\frac{2xy}{r^{4}}dydx
\end{equation}
 We  obtain the following
 asymptotical dreibein \cite{s7}
 $$e^{0}_{0}=\frac{1}{\sqrt{a}}, \enspace
e^{0}_{1}=\sqrt{a}(1+\frac{\omega
_{0}}{a})\frac{y}{r^{2}}$$
\begin{equation}
 e^{0}_{2}=-\sqrt{a}(1+\frac{\omega_{0}}{a})\frac{
x}{r^{2}}, \enspace e^{1}_{1}
=1+\frac{ay^{2}}{2r^{4}}
\end{equation}
 $$ e^{1}_{2}=\frac{axy}{r^{4}}, \enspace
e^{2}_{2}=1+\frac{ax^{2}}{2r^{4}}$$
Hence it can be shown that
$$\lim_{r\rightarrow \infty}r\omega_{\mu ab}=0$$
Thus
$$\int_{\partial\Sigma}\omega_{\mu ab}dx^{\mu}=0$$
So the total energy vanishes also. In the limit,
$\mu\rightarrow\infty$, eq.(50) has a solution with
dreibein and spin connection\cite{s15}
$$e^{0}=dt+\frac{\kappa J}{2\pi r^{2}}{\bf r}\times d{\bf r}
\,\,\,\,\,\,\,
{\bf e}=(1-\frac{\kappa m}{2\pi})d{\bf r}+\frac{\kappa m}
{2\pi r^{2}}{\bf r}
({\bf r}\cdot d{\bf r})$$
\beq
\omega^{0}=\frac{\kappa m}{2\pi r^{2}}{\bf r}\times d{\bf r}
\,\,\,\,\,\,\,
\omega^{i}=0
\eeq
we have
\beq
P_{a}=(m,0,0)
\eeq
which is the same as in\cite{s16}.
\section*{V. Discussions}
\indent As an end, we make some discussions. First, general
covariance is a
fundamental demand for conservation laws in general relativity.
our definition
eq.(39) (40) of energy-momentum is coordinate independent. As
the definition
of angular-momentum\cite{s7}, under local SO(1,2) transform
$e^{a}\rightarrow
\Lambda^{a}\,_{b}(x)e^{b}$, where $\Lambda^{a}\,_{b\mid\p\Sigma}
=L^{a}\,_{b}
=const.$, we have $P^{a}\rightarrow L^{a}\,_{b}P^{b}$. Second,
it is worth
while noting that for Deser's solution, the source stress-energy
tensor
of which is given {\it a priori}(the energy-momentum vanishes
while Deser's
gravitational mass vanishes only when $m+\sigma\mu=0$). This is
quite
different from the solution eq.(57) in the limit
$\mu\rightarrow\infty$.
The reason is that, though the form of eq.(51) agrees with eq.(57),
the
fall-off is substantially different. Remember that in Deser's
solution,
the metric is linearized $g_{\mu\nu}=\eta_{\mu\nu}+h_{\mu\nu}$,
which
is a good approximation on condition that $\mid h_{\mu\nu}\mid\ll 1$.
Yet in Deser's
solution, $h_{ij}=\phi\delta_{ij}$ and $\phi\sim \ln r$, so
$h_{\mu\nu}$ does
not satisfy the confition. We expect a solution with the same
$T^{\mu\nu}$ as Deser's while without the difficulty.

\vskip 0.3in
\underline{\bf Acknowledgement} S.S. Feng is indebted to Prof. S.
Randjbar-Daemi for his invitation for working at ICTP for three
months.This work is supported by the National Science
Foundation of China under Grant No. 19805004 and the Funds for Young Teachers
of Shanghai Education Council.

\end{document}